\documentclass[]{elsart}

\usepackage{epsfig}
\usepackage{graphics}
\usepackage{graphicx}
\usepackage{url}
\usepackage{subfigure}

\begin{document}

\begin{frontmatter}

\title{Towards \emph{Physarum} Binary Adders}

\author{Jeff Jones and Andrew Adamatzky} 

\address{Unconventional Computing Centre, University of the West of England, Bristol, United Kingdom}

\date{\today}

\begin{abstract}
Plasmodium of \emph{Physarum polycephalum} is a single cell visible by unaided eye. 
The plasmodium's foraging behaviour is interpreted in terms of computation. Input 
data is a configuration of nutrients, result of computation is a network of plasmodium's 
cytoplasmic tubes spanning sources of nutrients. Tsuda et al (2004) experimentally 
demonstrated that basic logical gates can be implemented in foraging behaviour of 
the plasmodium. We simplify the original designs of the gates and show --- in computer models --- that
the plasmodium is capable for computation of two-input two-output gate 
$\langle  x, y \rangle \rightarrow \langle  xy, x+y \rangle$
and  three-input two-output $\langle  x, y, z \rangle \rightarrow \langle  \overline{x}yz, x+y+z \rangle$. 
We assemble the gates in a binary one-bit adder and demonstrate validity of the design using computer simulation. 

\vspace{0.5cm}


\end{abstract}


\maketitle

\end{frontmatter}

\section{Introduction}

A plasmodium is a vegetative state of acellular slime mould \emph{Physarum polycephalum}. The plasmodium feeds on microscopic food particles, including microbial life forms. The plasmodium placed in an environment with distributed nutrients  develops a network of protoplasmic tubes spanning the nutrients' sources. Te topology of the  plasmodium's protoplasmic network optimizes the plasmodium's harvesting on the scattered sources of nutrients and makes more efficient flow and transport of intra-cellular components~\cite{nakagaki_2000,nakagaki_2001,nakagaki_2001a,nakagaki_iima_2007}.
 
The plasmodium is capable for approximation of shortest path~\cite{nakagaki_2001a}, computation of planar proximity 
graphs~\cite{adamatzky_toussaint} and plane tessellations~\cite{shirakawa}, primitive memory~\cite{saigusa}, basic logical computing~\cite{tsuda_2004}, and control of robot navigation~\cite{tsuda_2007}. The plasmodium can be considered as a 
general-purpose computer because the plasmodium simulates Kolmogorov-Uspenskii machine --- the storage modification machine operating on a colored set of graph nodes~\cite{adamatzky_ppl_2007}.  

The paper is structured as follows. In Sect.~\ref{physarumgates} we introduce the experimental gates invented in~\cite{tsuda_2004}
and re-interpret the gates as multi-output logical gates. We analyse asynchronism and reversibility of the gates in 
Sects.~\ref{asynchronism} and \ref{reversinggradients}. We simulate the gates in a particle-swarm model in Sect.~\ref{modellinggates}.  We assemble the gates in the one-bit half-adder and simulate the adder's behaviour in Sect.~\ref{modellingadder}.

\section{\emph{Physarum} gates}
\label{physarumgates}

\emph{Physarum} gates constructed in~\cite{tsuda_2004} were made of agar gel channels. Presence of a plasmodium in an 
input channel represents logical input {\sc Truth} ('1') and absence of plasmodium --- logical input {\sc False} ('0').
Values of signal in output channels are encoded similarly. Sources of chemo-attractants (glucose) are placed near
exits of output channels. The chemo-attractants diffusing in the agar gel channels establish gradients which guide the plasmodia
towards closest sources of attractants. 

In experiments discussed in~\cite{tsuda_2004} plasmodia inoculated in different input channels exhibited an aversion toward each other. They did not merge. If propagating plasmodium $p_1$ 
encountered another plasmodium $p_2$ in a channel $p_1$ wanted to travel in the plasmodium $p_1$ chosen another route of propagation. The fact that two `colliding' plasmodia do not merge was also supported by our experiments on constructing Voronoi 
diagram by plasmodia inoculated on nutrient-rich agar~\cite{adamatzky_toussaint,shirakawa}. Approaching wave-fronts of growing plasmodia usually `freeze' for an up 16 hours, when collide, however later the fronts merge. Outcomes of interaction between
two localized (i.e. propagating as wave-fragments) plasmodia depends on many factors, and 'elastic' collision is just one amongst many scenarios of the plasmodia interactions.

In paper~\cite{tsuda_2004} some output channels of \emph{Physarum} gates were considered as buffers. Let us 
now slightly redesign the gates~\cite{tsuda_2004} and interpret all outputs of the gates as Boolean logic values.

\begin{figure}
\centering
\subfigure[]{\includegraphics[scale=0.3]{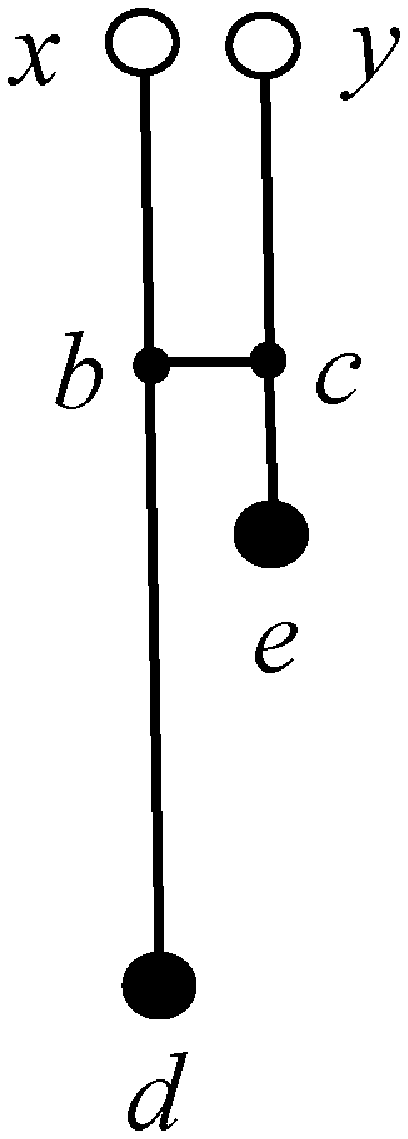}}
\subfigure[]{\includegraphics[scale=0.3]{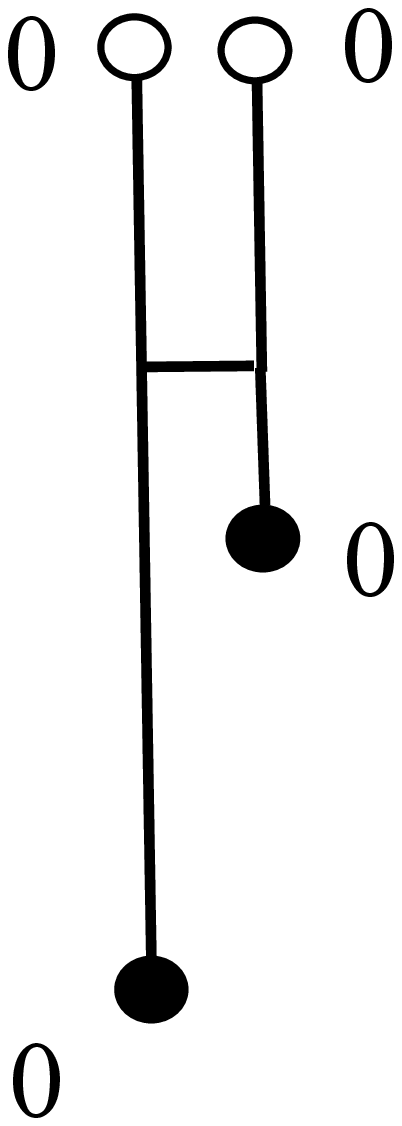}}
\subfigure[]{\includegraphics[scale=0.3]{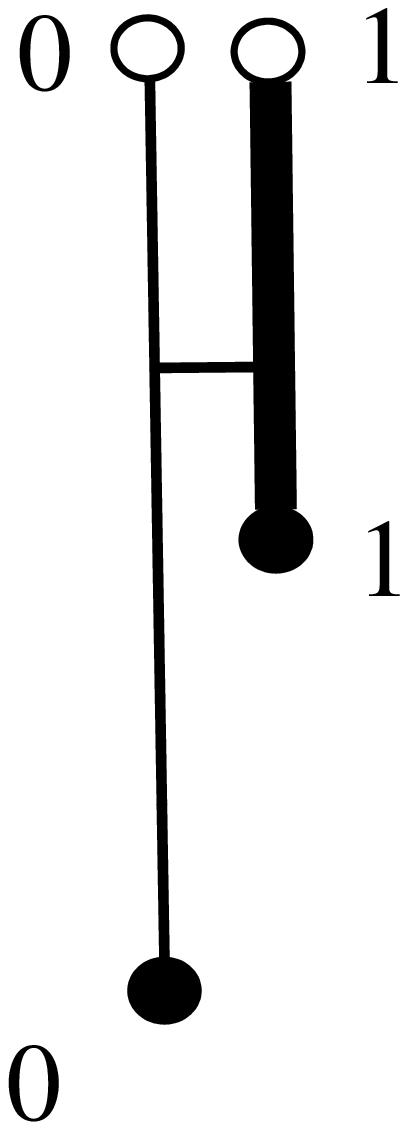}}
\subfigure[]{\includegraphics[scale=0.3]{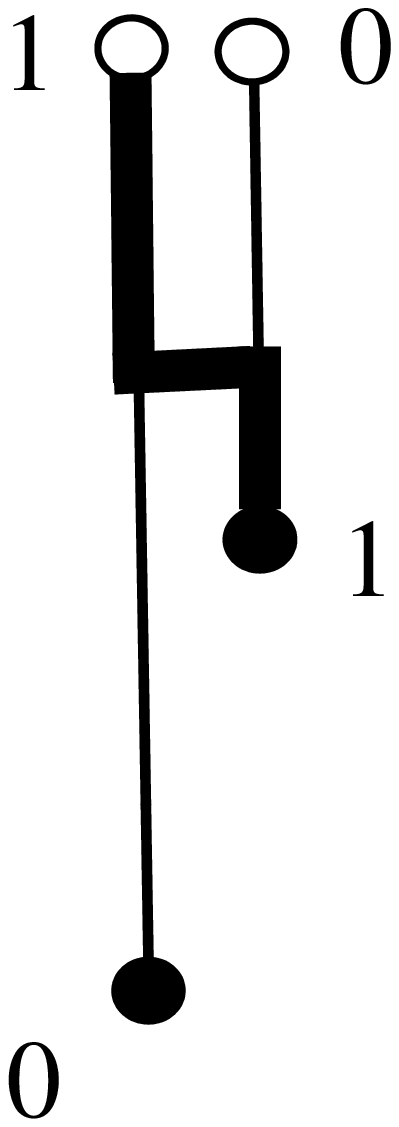}}
\subfigure[]{\includegraphics[scale=0.3]{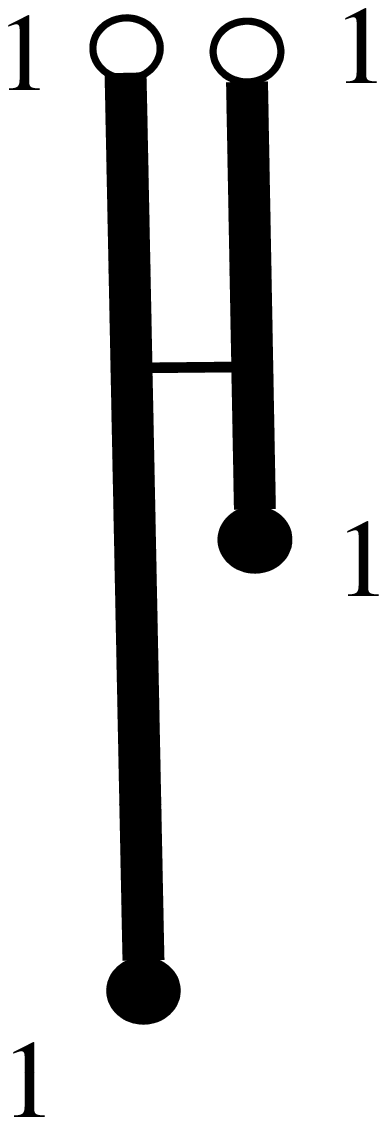}}
\subfigure[]{\includegraphics[scale=0.3]{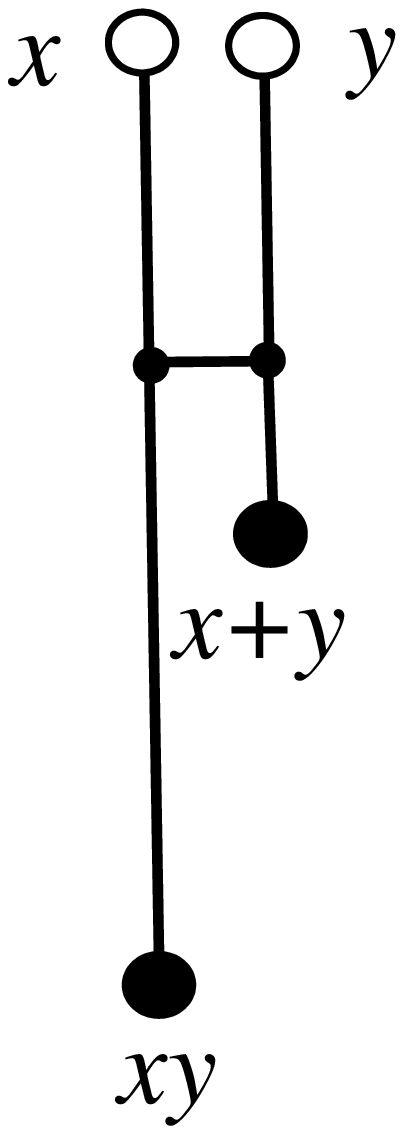}}
\caption{Scheme of $G_1$ gate: (a)~landmark points are shown; (b)--(e)~configuration of plasmodia in gates for all combinations of input values --- $x=0, y=0$~(b),  $x=0, y=1$~(c),  $x=1, y=0$~(d),  $x=1, y=1$~(e), 
the plasmodia bodies are shown by thick lines; (f)~input-output logical function realized by the gate. Chemo-attractants are placed in sites marked by solid black discs. }
\label{AND_OR}
\end{figure}

Consider $G_1$ gate in (Fig~\ref{AND_OR}a). Physical structure of the gate satisfies the following
constraints $|xb|=|yc|$ and $|bd| > |bc| + |ce|$ (Fig~\ref{AND_OR}a).  Chemo-attractants are placed in sites $d$ and $e$. 
We assume strength of attraction to $d$ ($e$) at point $p$ is proportional to distance $|pd|$ ($|pe|$) 
(Fig~\ref{AND_OR}a).  

Situations corresponding to input values $(0,0)$, $(0,1)$ and $(1,0)$ are simple. 
When no plasmodia are inoculated in $x$ and $y$ nothing appears at outputs $d$ and $e$ (Fig~\ref{AND_OR}b). 
When plasmodium is placed only in site $y$ the plasmodium follows the route $(yc)(ce)$ (Fig~\ref{AND_OR}c). 
If plasmodium inoculated only in site $x$ the plasmodium follows the route $(xb)(bc)(ce)$ (Fig~\ref{AND_OR}d).

The main trick of the gate is in how input values $x=1$ and $y=1$ are handled. The
plasmodia are inoculated in sites $x$ and $y$ (Fig~\ref{AND_OR}d). The plasmodium 
growing from site $y$ follows route $(yc)(ce)$. The plasmodium growing from site $x$ 
tends to follow route $(xb)(bc)(ce)$, however part of the route $(ce)$ is already 
occupied by another plasmodium. Therefore the plasmodium, starting in $x$, grows 
along the route $(xb)(bd)$~(Fig~\ref{AND_OR}d).

A table of transformation $\langle  x, y \rangle \rightarrow \langle  d, e \rangle$ shows that the gate $G_1$ (Fig~\ref{AND_OR}f)
implements logical conjunction and logical disjunctions $\langle  x, y \rangle \rightarrow \langle  xy, x+y \rangle$ at 
the same time but on two different outputs.

\begin{figure}
\centering
\subfigure[]{\includegraphics[scale=0.3]{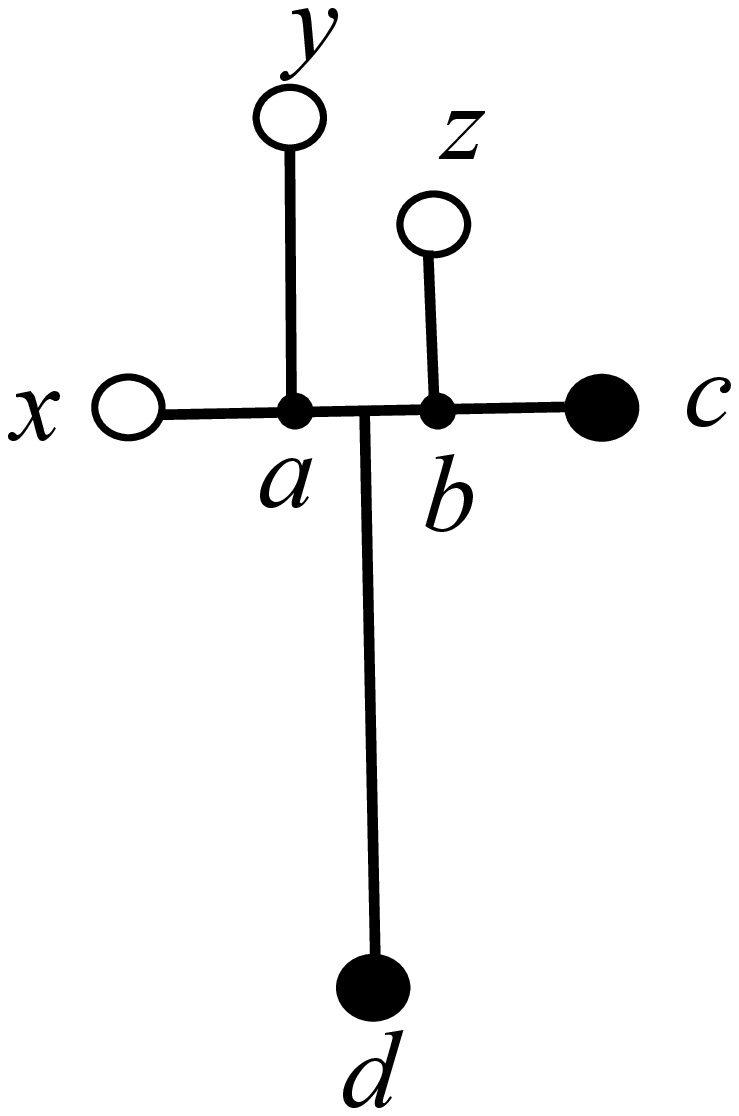}}
\subfigure[]{\includegraphics[scale=0.3]{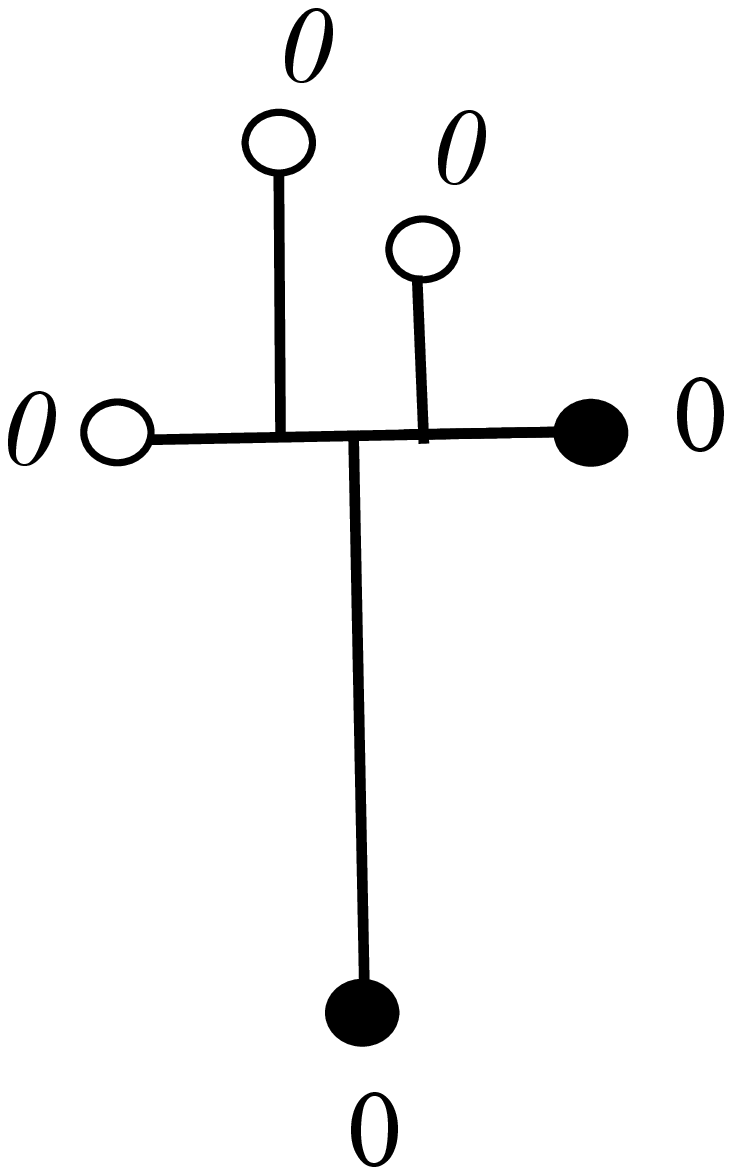}} 
\subfigure[]{\includegraphics[scale=0.3]{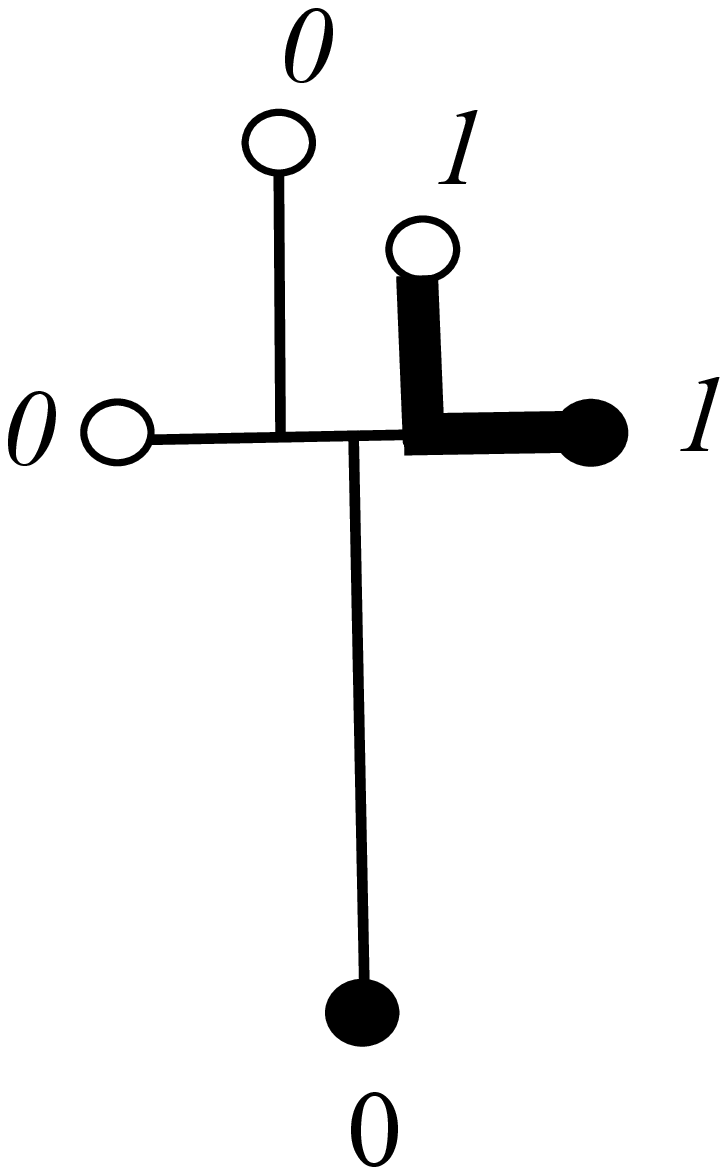}} 
\subfigure[]{\includegraphics[scale=0.3]{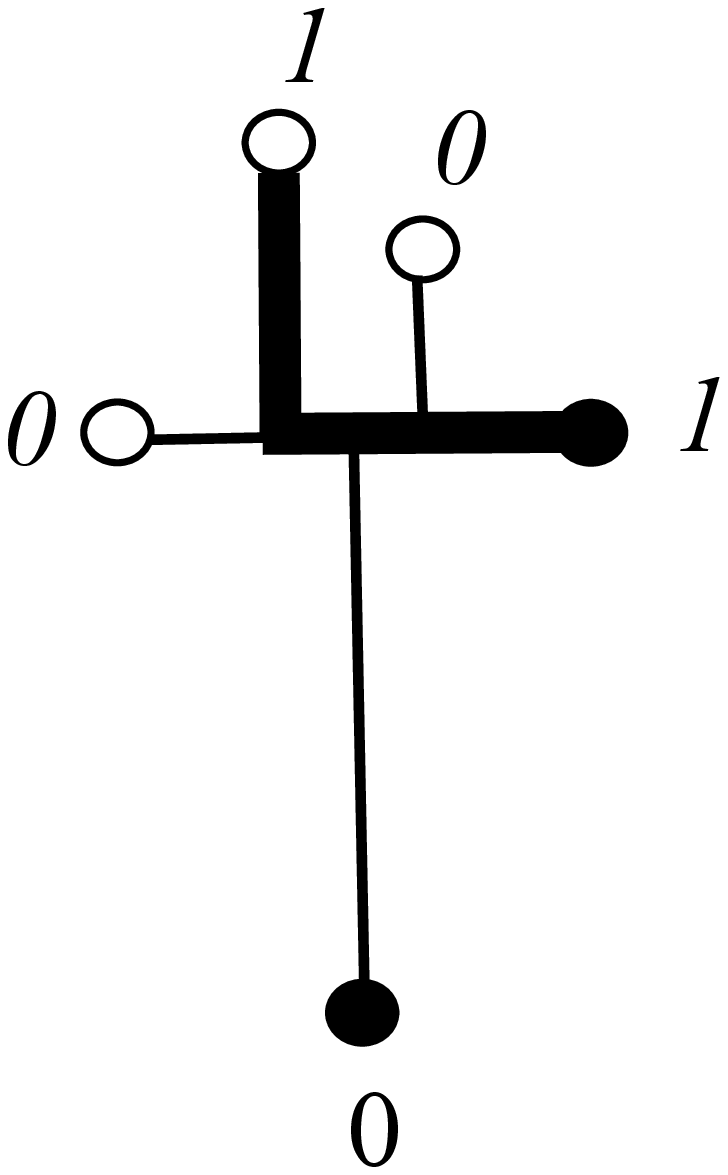}} 
\subfigure[]{\includegraphics[scale=0.3]{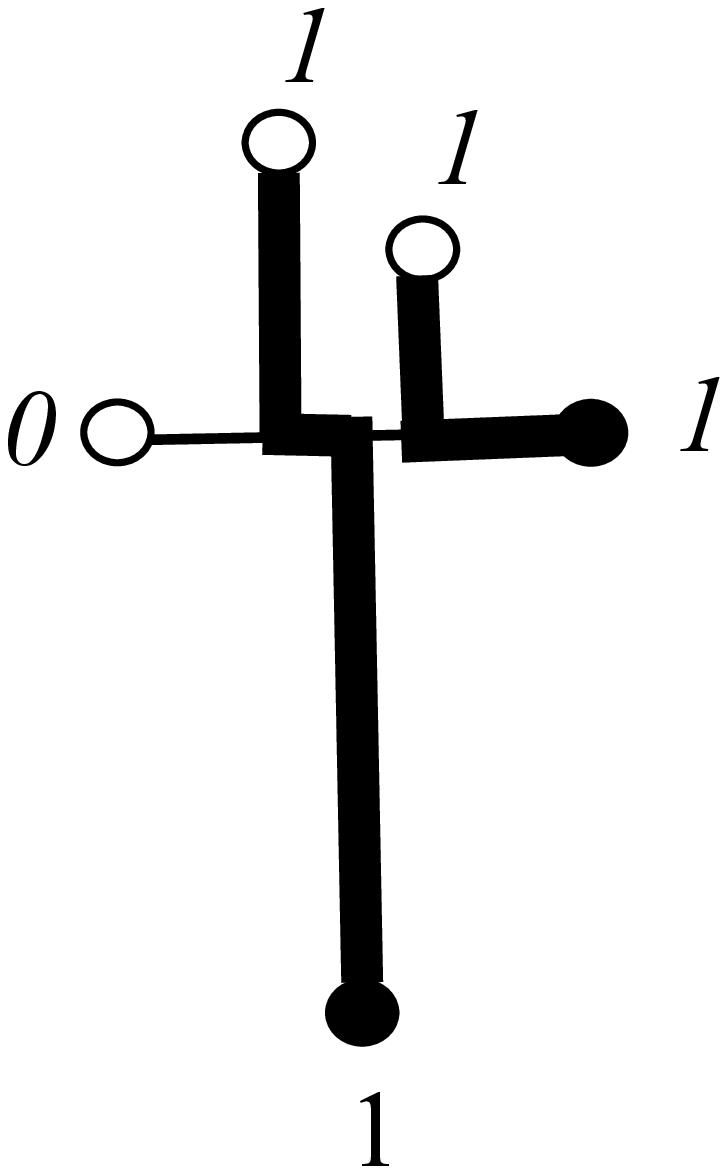}} \\
\subfigure[]{\includegraphics[scale=0.3]{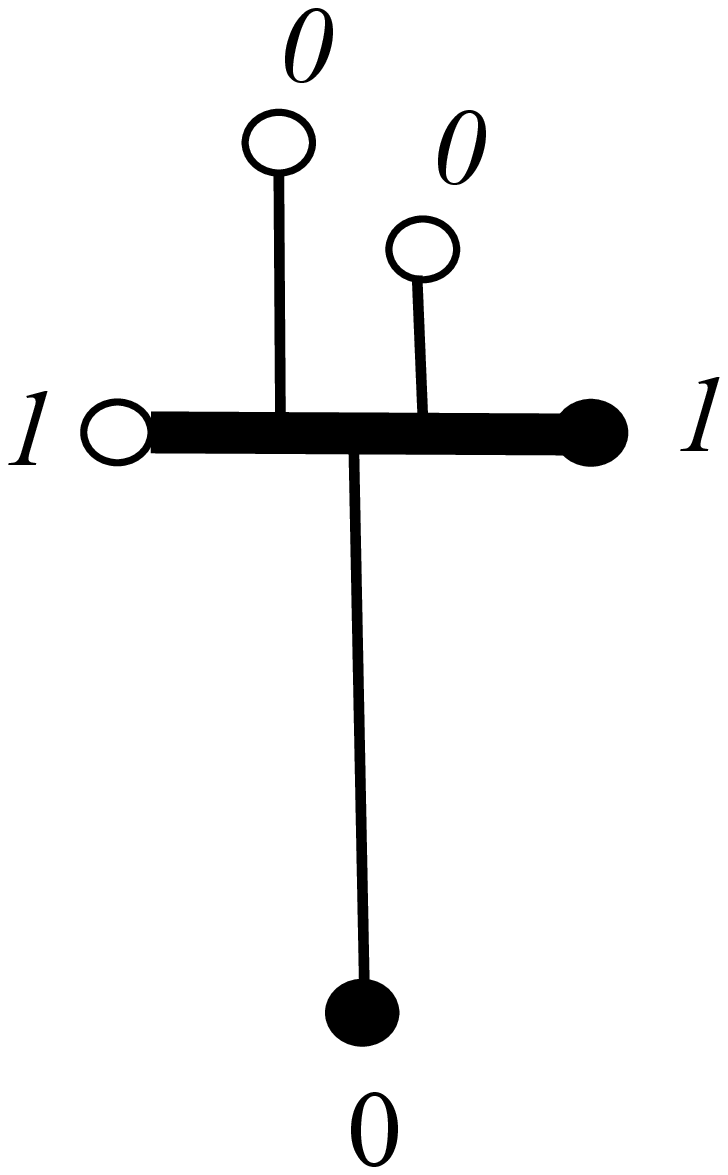}} 
\subfigure[]{\includegraphics[scale=0.3]{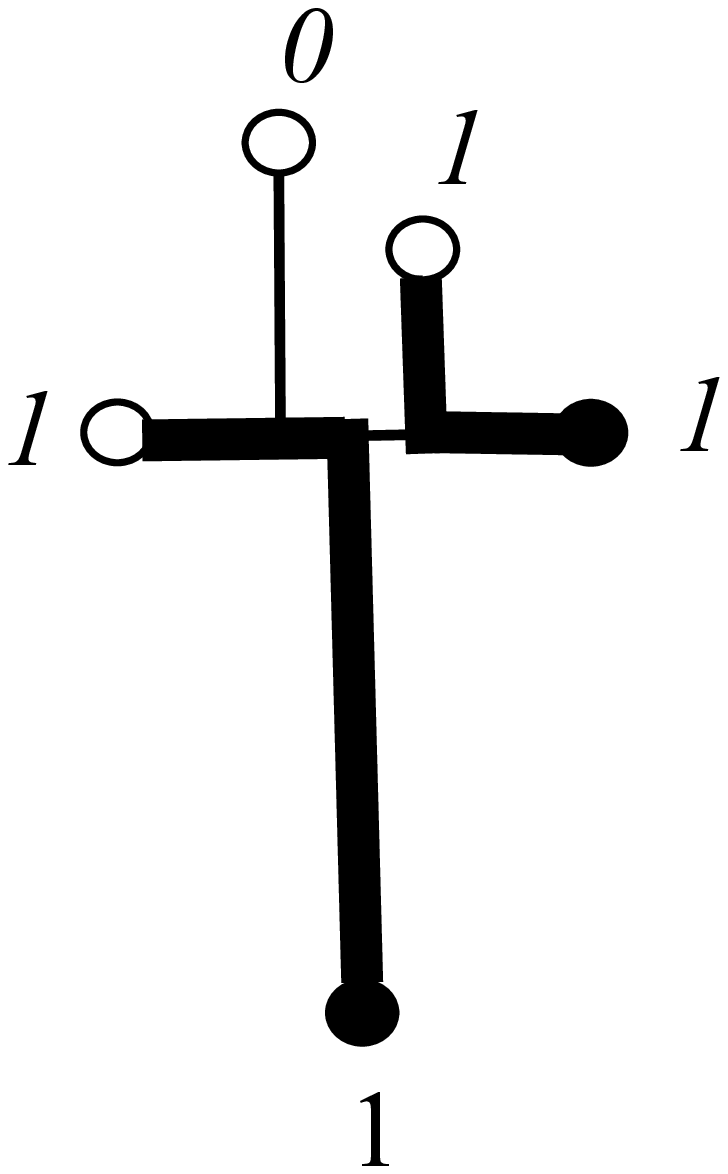}} 
\subfigure[]{\includegraphics[scale=0.3]{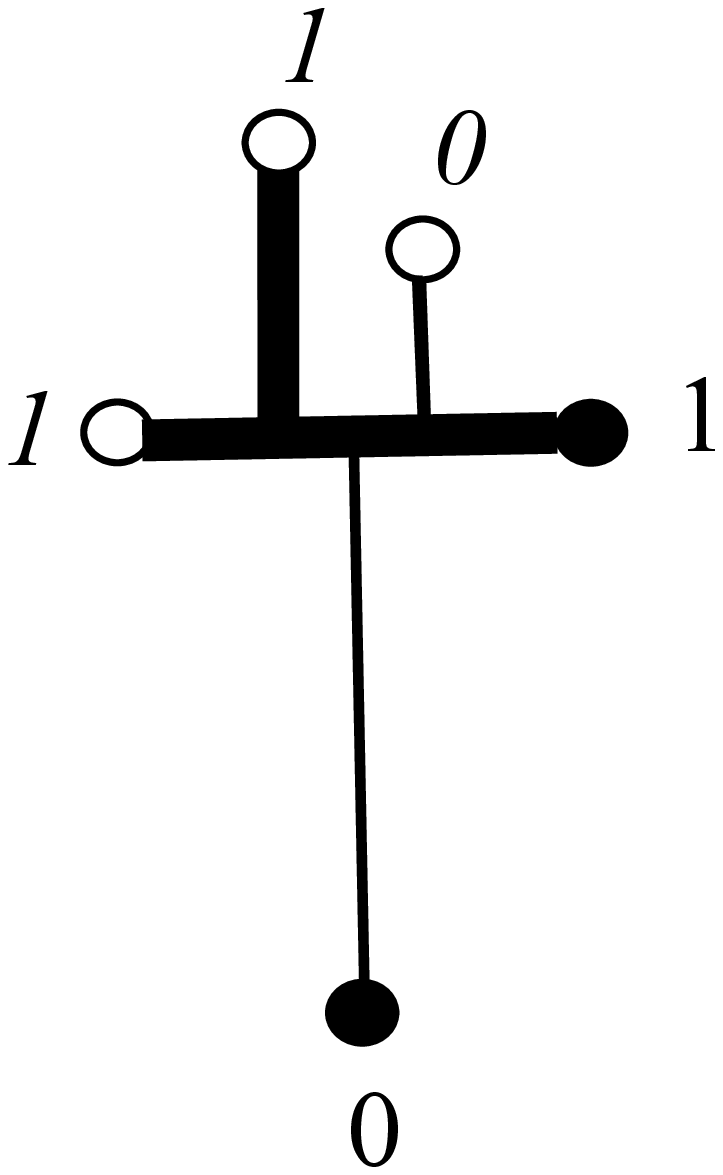}} 
\subfigure[]{\includegraphics[scale=0.3]{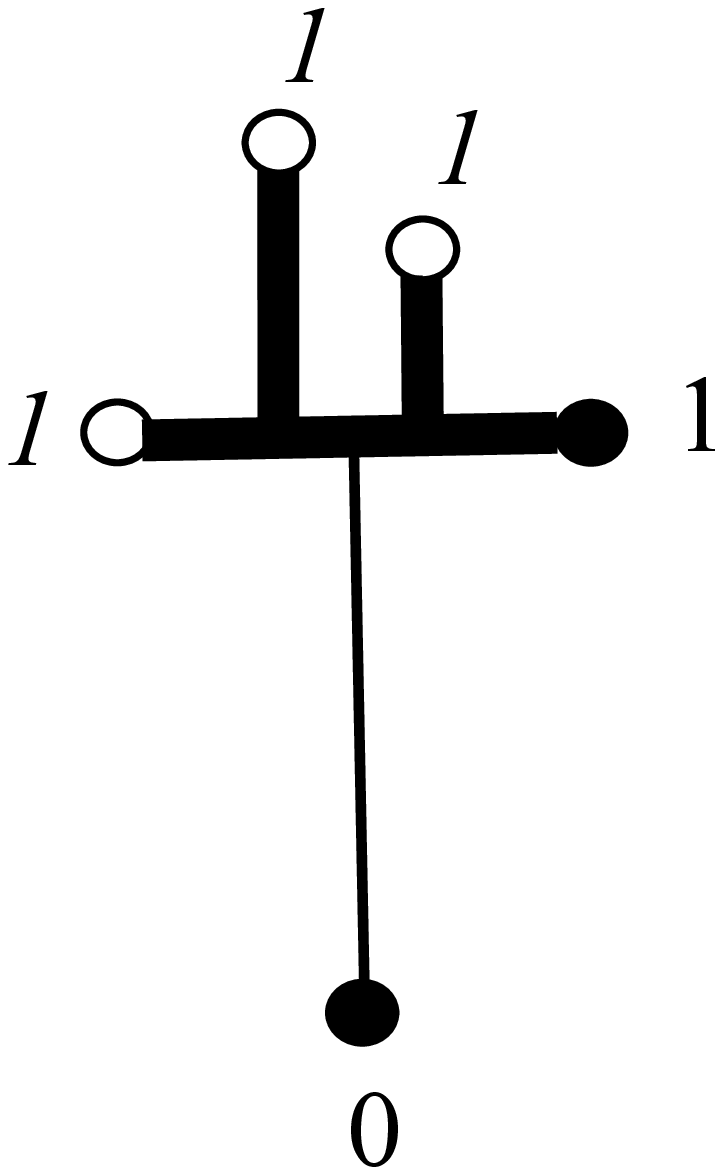}} 
\subfigure[]{\includegraphics[scale=0.3]{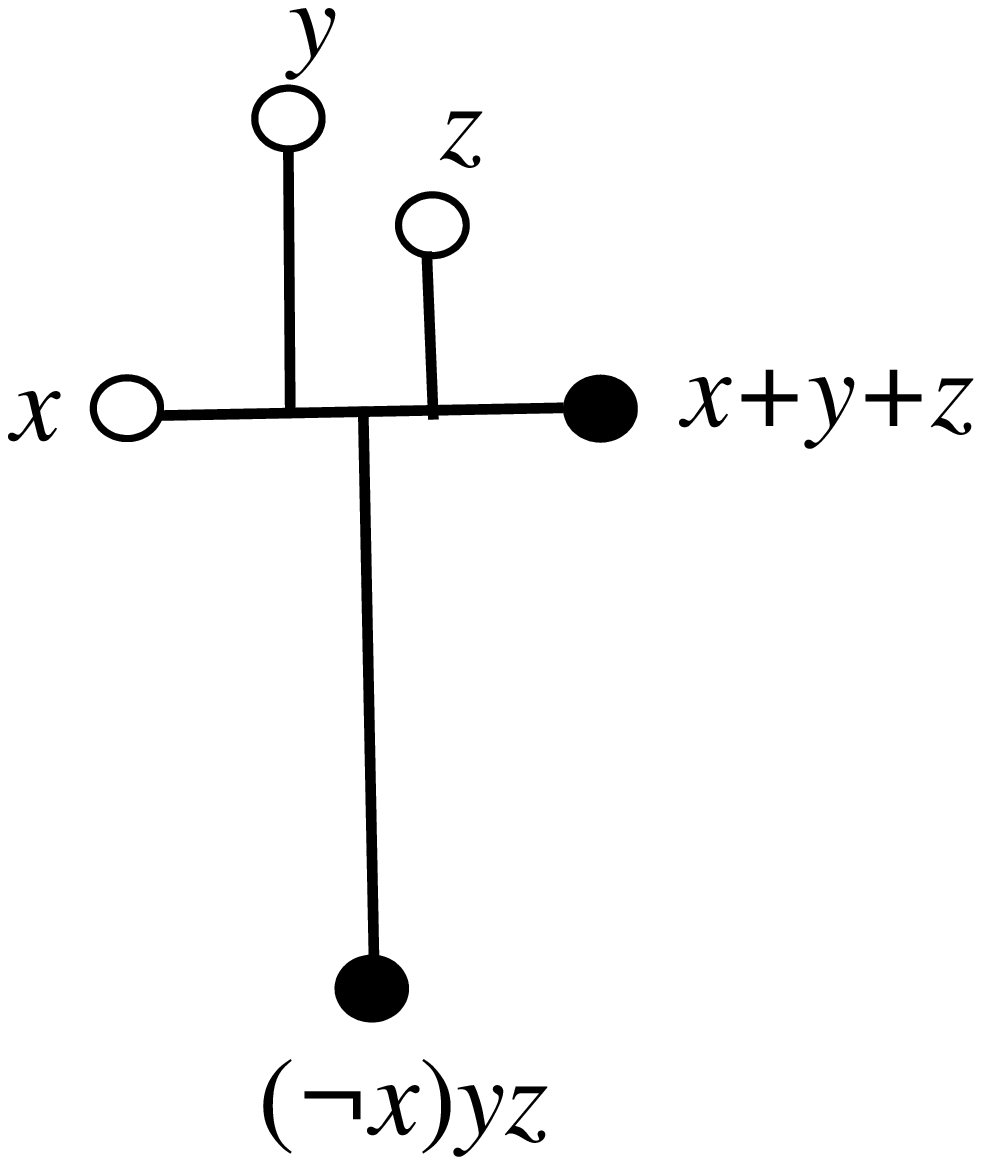}} 
\caption{Scheme of $G_2$ gate.
(a)~landmark points are shown; 
(b)--(i)~configuration of plasmodia in gates for all values of input tuple $\langle  x, y, z \rangle$:
(b)~$\langle  000 \rangle$, (c)~$\langle  001 \rangle$, (d)~$\langle  010 \rangle$,
(e)~$\langle  011 \rangle$, (f)~$\langle  100 \rangle$, (g)~$\langle  101 \rangle$,
(h)~$\langle  110 \rangle$, (i)~$\langle  111 \rangle$, the plasmodia bodies are shown by thick lines;
(j)~input-output logical function realized by the gate. Input are marked with circles, outputs with solid discs. Chemo-attractants are placed in sites marked 
by solid black discs.}
\label{NOT}
\end{figure}

Geometrical structure of $G_2$  gate is shown in Fig.~\ref{NOT}. Chemo-attractants are placed in 
sites $c$ and $d$ and plasmodia can be inoculated in sites $x$, $y$ and $z$ (Fig.~\ref{NOT}a).
Lengths of channels in the gate satisfy the following conditions:  $|xc| < |xd|$, $|ac| < |ad|$, 
$|bc| < |bd|$, and $|zb| + |bc| < |ya| + |ac|$. 

In~\cite{tsuda_2004} input channels $y$ and $z$ (Fig.~\ref{NOT}a) were assigned to constant {\sc Truth} inputs 
an output  channel $c$ to a buffer (unused output to collect `excess' of plasmodium). Let 
consider  scenario when all three input can take values `0' and '1' and both outputs have 
a meaning. 

If plasmodium placed in site $z$ it propagates toward closest attractant-site $c$ (Fig.~\ref{NOT}c); 
similarly a plasmodium inoculated in site $y$  propagates towards attractant-site $c$ (Fig.~\ref{NOT}d).
When plasmodia are placed in sites $y$ and $z$ simultaneously, the plasmodium from the site $z$ follows 
the route $(zb)(bc)$ and thus blocks the way for plasmodium propagating from $y$ (Fig.~\ref{NOT}e)). 
Therefore the plasmodium originating in $y$ moves to attractant-site $d$ (Fig.~\ref{NOT}e). 
The situations sketched in Fig.~\ref{NOT}g--j can be described similarly. Considering the transformations
$\langle  x, y, \rangle \rightarrow \langle  c, d \rangle$ we find that the gate implements the following
logical function $\langle  x, y \rangle \rightarrow \langle  x, x\overline{y} \rangle$. If  $y$- and $z$-iputs 
are constant {\sc Truth}, $y=1$ and $z=1$, the gate $G_2$ is a negation (this how it was initially designed 
in~\cite{tsuda_2004}).

\begin{figure}
\centering
\includegraphics[scale=0.3]{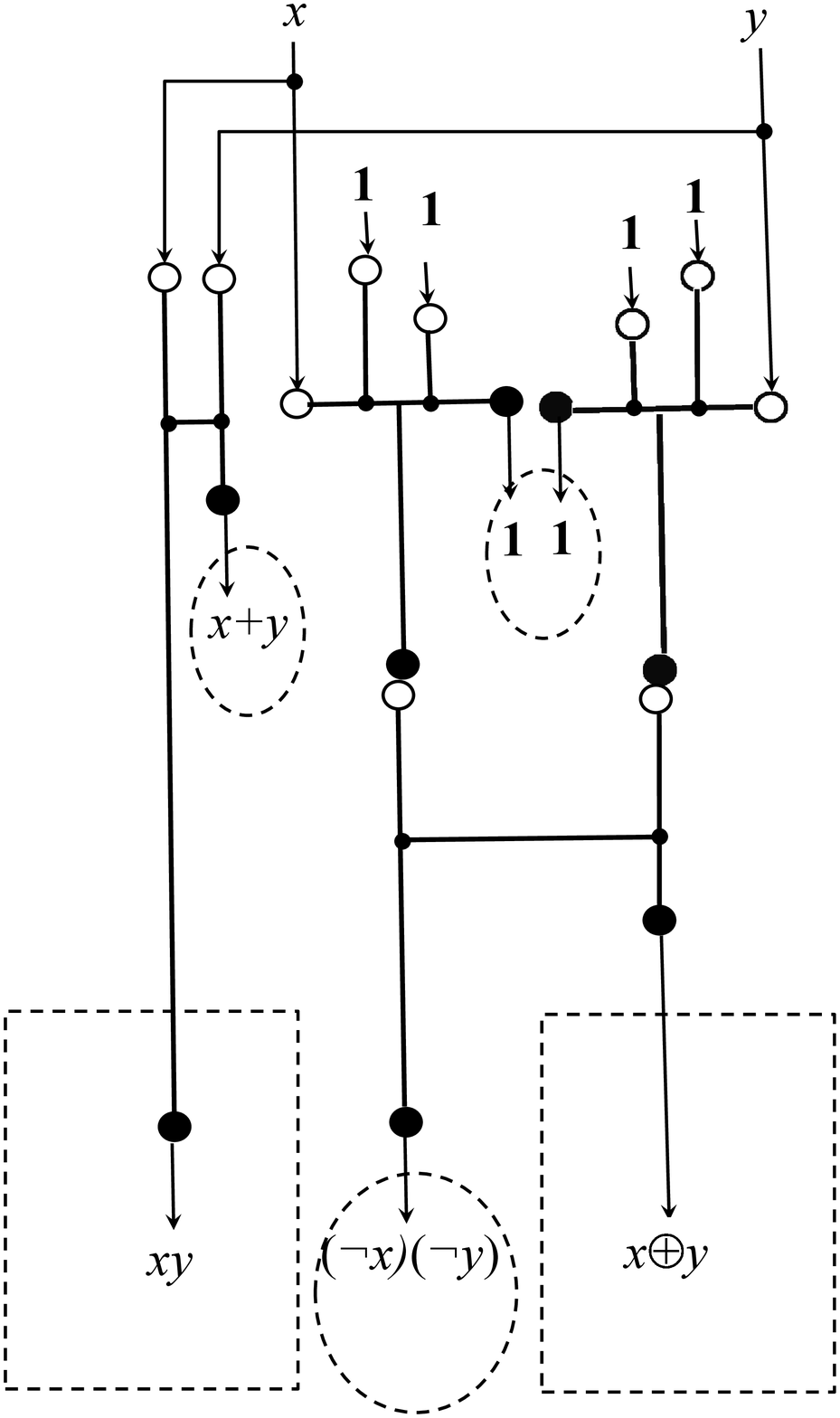}
\caption{Scheme of {\emph Physarum} one-bit half-adder. Input variables are $x$ and $y$, {\bf 1} on
input channels represent constant {\sc Truth}. Carry value $xy$ and sum $x \oplus y$ are 
highlighted by dotted rectangle, unused outputs $x + y$, $\mathbf 1$ and $(\neg x)(\neg y)$ by dotted ellipses.}
\label{halfadder}
\end{figure}

\emph{Physarum} gates $G_1$ and $G_2$ can be cascaded by linking output gel-channels of one gate to input 
gel-channels of another gate. An example of such cascading in a form of one-bit half adder is shown in 
Fig.~\ref{halfadder}. Four pieces of plasmodium are fed in input channels as constant {\sc Truth}. The plasmodia 
representing Boolean variables $x$ and $y$ are multiplied or branched and fed into gate $G_1$ and two copies of 
gate $G_2$. Output channels of gates $G_2$ are fed into data channels of another gate $G_1$. In addition to 
results we are looking for --- $xy$ and $x \oplus y$ --- the circuit (Fig.~\ref{halfadder}) produces several 
byproducts: $x+y$, $(\neg x)(\neg y)$ and two copies of constants {\sc Truth}. These signals can be used further down in the 
chain of computation or routed in the buffer zones (plasmodium pool). Plasmodia representing constant {\sc Truth} can 
be also rerouted back to control inputs of gates $G_2$.

\section{Asynchronism}
\label{asynchronism}

Synchronization of signals is amongst key factors in proper functioning of logical circuits. Architecture of 
\emph{Physarum} gates allow for a certain degree of asynchronism. Let us evaluate a degree of asynchronism of 
gates $G_1$ and $G_2$.

Assume outputs of gate $G_1$ are read in $w \geq|xb|+|bc|+|bd|$ time units after on the plasmodia entered its data channel.
Let plasmodia representing logical variables $x$ and $y$ enter their channels at time steps $\tau_x$ and $\tau_y$.
Let $x=1$ and $y=1$. Then $x$-plasmodium must reach site $c$ of gate $G_1$ (Fig.~\ref{AND_OR}a) after the site 
$c$ is occupied by $y$-plasmodium. That is $\tau_x + |xb| \geq \tau_y + |yc|$. Due to $|xb|=|yc|$ (Fig.~\ref{AND_OR}a)
we have $\tau_x \geq \tau_y$. If $x=1$ the $x$-plasmodium must reach exit of the output channel $(xd)$ before `signal reading' 
time-window closed. Therefore we have $\tau_y \leq \tau_x \leq \alpha_1 = w - |xd|+|bc|$. The parameter $\alpha_1$ is a degree of
asynchronism of gate $G_1$. In the same manner we obtain a constraint on timing $\tau_x$, $\tau_y$ and $\tau_z$ of signals 
$x$, $y$ and $z$ in gate $G_2$: $\tau_z \leq \tau_y \leq \tau_x \leq w - |ya| - |ad|$. Thus degree of asynchronism of gate 
$G_2$ is $\alpha_2 = w - |ya| - |ad|$.

Architectures of  \emph{Physarum} gates $G_2$ (Fig.~\ref{NOT}) and 
$G_1$ (Fig.~\ref{AND_OR}) assume gradients of chemo-attractants from 
output sites (where sources of attractants) are placed to input sites (solid 
black discs in Figs.~\ref{NOT} and~\ref{AND_OR}). What will happen if we reverse the
gradients and place sources of chemo-attractants in input sites of original gates 
and consider output sites of original gates as inputs of new gates (Fig.~\ref{reversedgates})?
We will write gates $G_1$ and $G_2$ with reversed gradients and input-outputs as $\overline{G_1}$
and $\overline{G_2}$. 

\section{Outcomes of reversing gradients of chemo-attractants}
\label{reversinggradients}

\begin{figure}
\centering
\subfigure[]{\includegraphics[scale=0.3]{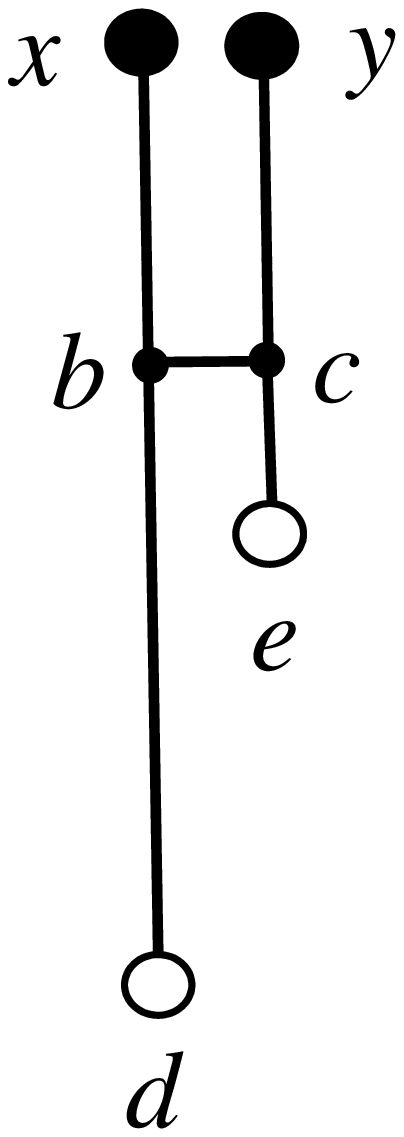}}
\subfigure[]{\includegraphics[scale=0.3]{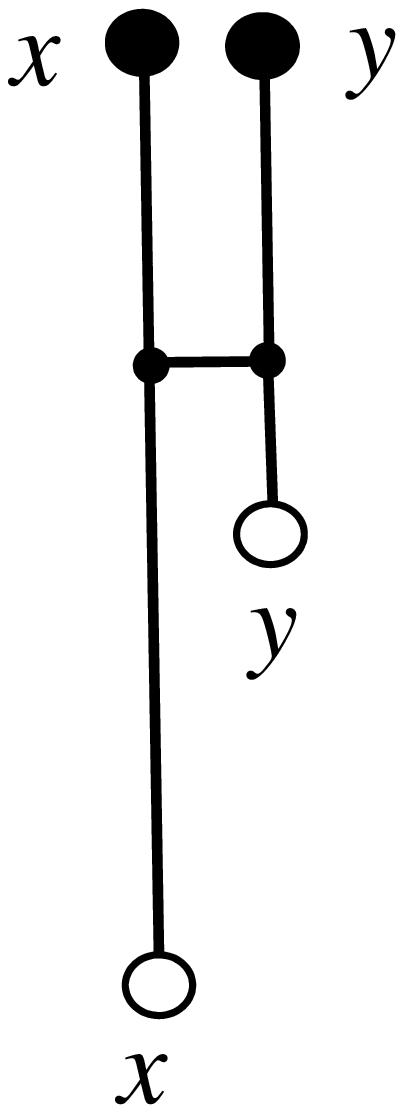}}
\subfigure[]{\includegraphics[scale=0.3]{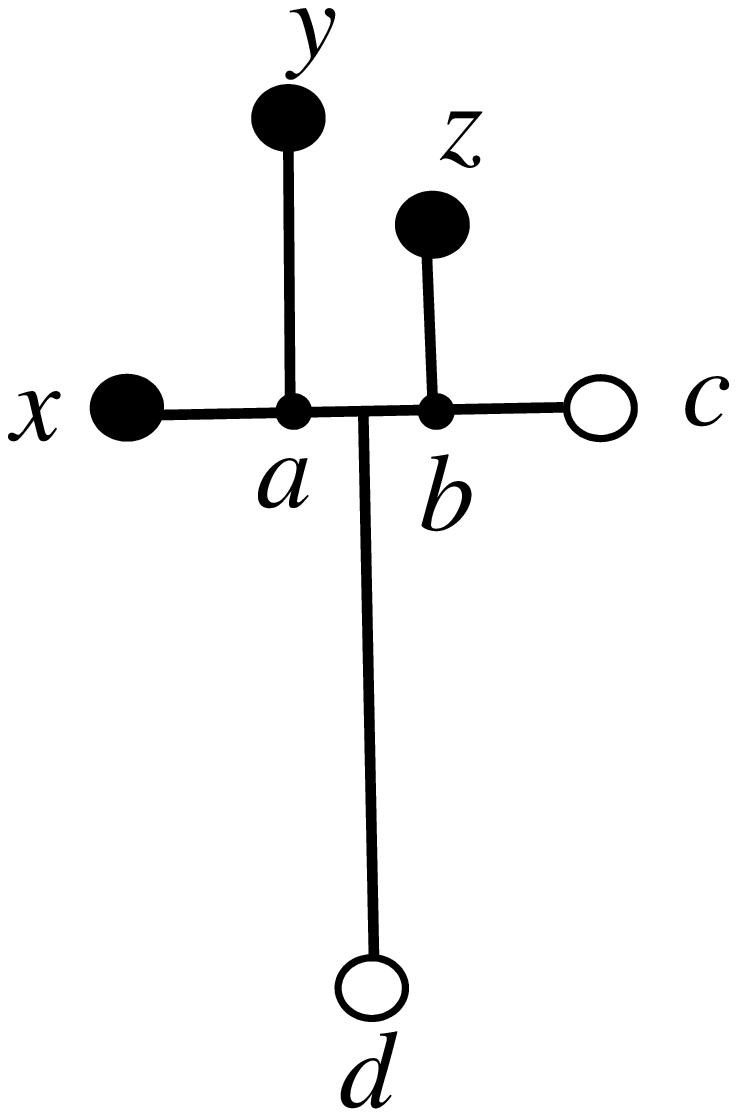}}
\subfigure[]{\includegraphics[scale=0.3]{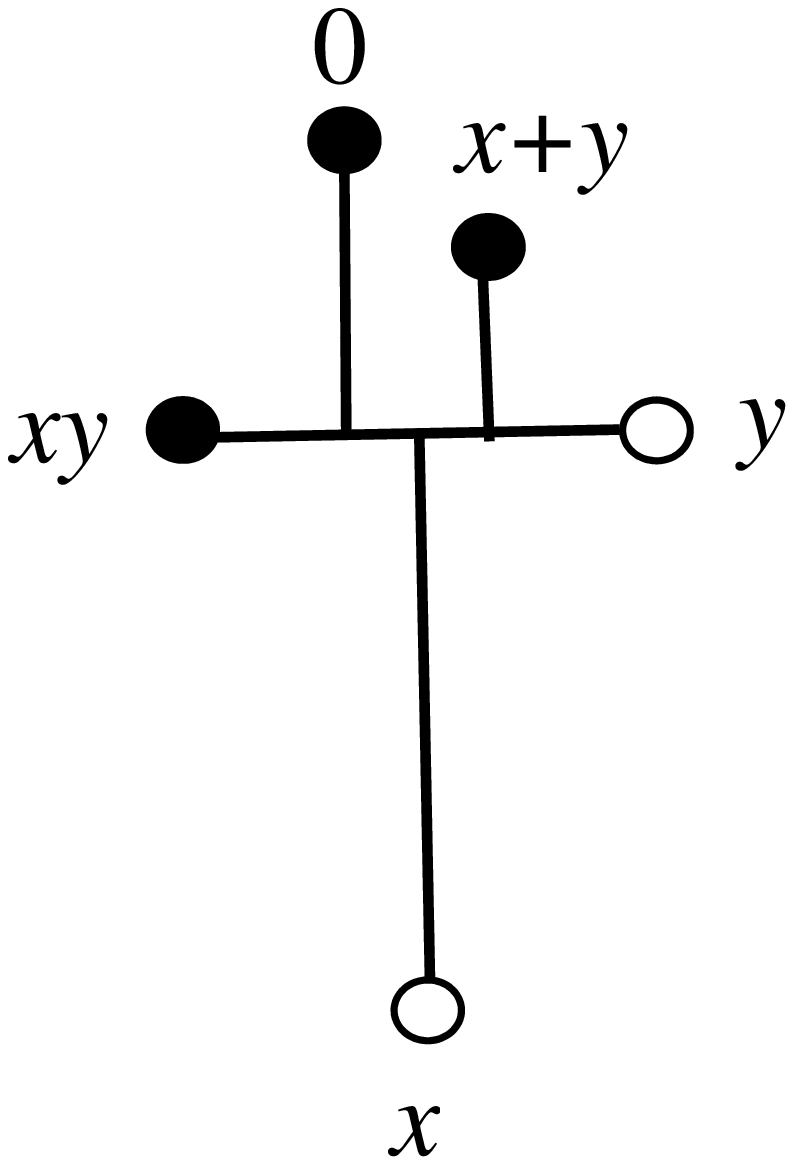}}
\caption{Gates $\overline{G_1}$(a)(b) and $\overline{G_2}$(c)(d). Landmarks are shown in (a) and (c), the gates
functional schemes in (b) and (d). Inputs are marked with circles, outputs with solid discs. 
Chemo-attractants are placed in sites marked by solid black discs.}
\label{reversedgates}
\end{figure}

Let us consider gate $\overline{G_1}$. If plasmodium placed in site $d$ (Fig.~\ref{reversedgates}a) it propagates towards
site $b$ (because it is the only choice) and then follows gradients towards closest source of chemo-attractants, site $x$.
Plasmodium inoculated in site $e$ travels along the route $(ec)(cy)$. Two plasmodium placed in sites $d$ and $e$ simultaneously,
do not interact. This shows that gate $\overline{G_1}: \langle  x, y \rangle \rightarrow \langle  x, y \rangle$ 
acts as a simple conductor of signals when polarity of chemo-attractant gradients is reversed (Fig.~\ref{reversedgates}b).

In gate $\overline{G_2}$ plasmodium placed in site $c$ always propagates towards closest source of attractants, site $z$. 
Plasmodium placed in site $x$ if $|xa| > |zb|$ but the plasmodium propagates to site $z$ if $|xa| < |zb|$ (Fig.~\ref{reversedgates}c). The plasmodium placed
in $c$ and/or $d$ never reaches site $y$, therefore output marked `$y$' is always '0'. Analysis of all combinations of input signals demonstrate that for $|xa| < |zb|$  we have $\overline{G_2}: \langle  x, y \rangle \rightarrow \langle  x, y \rangle$, 
and for $|xa| > |zb|$ we have $\overline{G_2}: \langle  x, y \rangle \rightarrow \langle  xy, x+y \rangle$ (Fig.~\ref{reversedgates}d).

In summary, when gradients of chemo-attractants and input-output swapped in gates $G_1$ and $G_2$ the gate $G_1$ becomes a simple conductor and the gate $G_2$ becomes gate $G_1$.

\section{Computational modelling of \emph{Physarum} gate behaviours}
\label{modellinggates}

 To model the \emph{Physarum} gate behaviours the three physical criteria identified in~\cite{tsuda_2004} and utilised in the design of the logic gates need to be implemented.  The criteria can be summarised as:
\\
1. \emph{Physarum} grows and moves towards nutrient chemoattractant gradients.\\
2. If two plasmodium fragments encounter each other, they will avoid contact where other routes exist.\\
3. If two plasmodium fragments cannot avoid contact, the plasmodia will fuse.\\

We employed the particle model of emergent transport network formation and evolution introduced in~\cite{jones_2010} to implement the gate behaviours, where a population of very simple mobile particles with chemotaxis-like sensory behaviour were used to construct and minimise spatially represented emergent transport networks in a diffusive environment. The particle approximation corresponds to a particle approximation of LALI (Local Activation Long-range Inhibition) reaction-diffusion pattern formation processes and exhibits a complex range of patterning by varying particle sensory parameters~\cite{jones_alife_2010}. We assume that each particle in the collective represents a hypothetical unit of \emph{Physarum} plasmodium gel/sol interaction which includes the effect of chemoattractant gradients on the plasmodium membrane (sensory behaviour) and the flow of protoplasmic sol within the plasmodium (motor behaviour). The summation of particle positions corresponds to a static snapshot of network structure whilst the collective movement of the particles in the network corresponds to protoplasmic flow within the network. 

Although the model is very simple in its assumptions and implementation it is capable of reproducing some of the spontaneous network formation, network foraging, oscillatory behaviour, bi-directional shuttle streaming, and network adaptation seen in \emph{Physarum}, using only simple, local functionality to generate the emergent behaviours. Details of the particle morphology, sensory and motor behavioural algorithms can be found in ~\cite{jones_uc09} and in this paper we use an extension of the basic model (without utilising oscillatory behaviour) to include plasmodium growth and adaptation (growth and shrinkage of the collective).

The environment is represented by a greyscale image where different values correspond to different environmental features (for example, habitable areas, inhabitable areas, nutrient sources). The particles move about their environment (a two-dimensional lattice) and sample sensory chemoattractant data from an isomorphic diffusion map. When particles move about their environment they deposit chemoattractant to the same structure. Chemoattractant gradients were represented by projection of chemoattractant to the diffusion map at the locations indicated on the gate schematic illustrations. The projection weight was set at 20 multiplied by the chemoattractant pixel value (255). The weight factor is high as chemoattractant is deemed to be completely absorbed when it encounters the edges of the chamber and a large weight value is necessary to ensure the required propagation distance. The diffusion kernel was a 7$\times$7 window for all experiments. Diffusion was achieved by the mean of the local window at each location in the diffusion map and damped at $10^{-4}$  (i.e. new value is equal to the mean multiplied by 1-$10^{-4}$). We assumed that diffusion of chemoattractant from a nutrient source was suppressed when the source was engulfed by particles. The suppression was implemented by checking each pixel of the food source and reducing the projection value (concentration of chemoattractants) by multiplying it by $10^{-3}$ if there was a particle within a 9$\times$9 neighbourhood surrounding the pixel. Particle sensor offset was 5 pixels, angle of rotation set to 45 degrees, and sensor angle was 45 degrees.

Growth and adaptation of the particle model population is currently implemented using a simple method based upon local measures of space availability (growth) and overcrowding (adaptation, or shrinkage, by population reduction). This is undoubtedly a gross simplification of the complex factors involved in growth and adaptation of the real organism (such as metabolic influences, nutrient concentration, waste concentration, slime capsule coverage, bacterial contamination etc.). However the simplification renders the population growth and adaptation more computationally tractable and the specific parameters governing growth and shrinkage are at least loosely based upon real environmental constraints of space and nutrient availability. Growth and shrinkage states are iterated separately for each particle and the results for each particle are indicated by tagging Boolean values to the particles. The growth and shrinkage tests were executed every three scheduler steps and the method employed is specified as follows. If there are 1 to 10 particles in a 9$\times$9 neighbourhood of a particle, and the particle has moved forwards successfully, the particle attempts to divide into two (i.e. a new particle is created) if there is an empty location in the immediate neighbourhood surrounding the particle. If there are 0 to 20 particles in a 5$\times$5 neighbourhood of a particle the particle survives, otherwise it is annihilated.

\subsection{Modelling individual gates}

\begin{figure}
\centering
\includegraphics[scale=0.8]{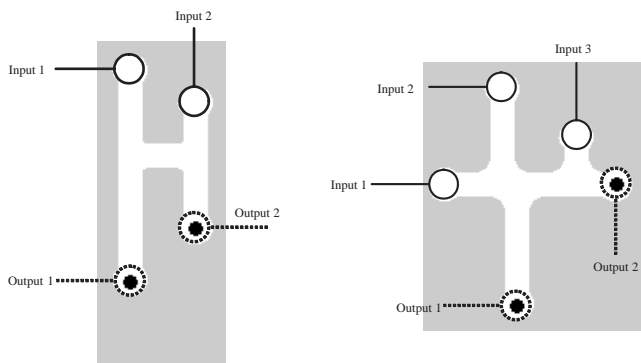}
\caption{Spatial implementation of logic gates $G_1$ and $G_2$ used in the particle model}
\label{jeff_g1_g2}
\end{figure}

To implement the gates using the model, the schematic illustrations in Fig.~\ref{AND_OR} and Fig.~\ref{NOT} were transformed into the spatial representations shown in Fig.~\ref{jeff_g1_g2}. The spatial pattern and greyscale encoding (boundaries, nutrient sources) is used to configure the diffusive map.

Particles were introduced (depending on logical input conditions) at the areas indicated by solid circles at the top of the gates. Strong sources of chemoattractant were introduced at the outputs indicated as enclosed by dashed circles. The chemoattractant diffused from the output locations along channels etched into the gate configurations (white areas) and chemoattractant was removed immediately on contact with boundaries of the channels (light grey areas). The particle population was inoculated at identical times at the inputs, sensing, growing and moving towards the propagating diffusion gradients. To `anchor' the growing paths to the start positions a very small amount of chemoattractant was also deposited at the respective start positions (the amount chosen was the lowest level needed to anchor the position without affecting the actual gate computation). Population inoculation and chemoattractant diffusion occurred at the same time and there was little or no directed growth of the population until the chemoattractants reached the source of inoculation.

The operation of the gates occurs due to the complex interactions between the chemoattractant diffusion gradients. Because there is a quantitative aspect to the chemoattractant gradient (i.e. particles sense not only the presence but also the strength of the diffusion), the gradient concentration is affected by the length and width of the gate channels~\cite{jones_passiveactive_2009}. The point at which the competing wave fronts meet is a spatial interface which delineates path choices in a similar way to those observed in chemical reaction-diffusion computations~\cite{steinbock_1995}. Thus, the environment is partially responsible for the initial selection of path choice. This `background processing' by the environment satisfies the first of the three aforementioned criteria for plasmodium gate construction.

Two more factors add to the complexity of gradient interactions: Firstly when the particle representation of the plasmodium engulfs a food source, the diffusion of chemoattractant from that source is suppressed (reduced by a factor of one thousand). This alters the strength of the gradient field from the engulfed source and the interface position where competing fronts meet shifts to reflect the new gradient field. Secondly, the collective movement of the particle population also results in local chemoattractant deposition along the path (this deposition is responsible for the local recruitment of particles by positive feedback and also acts to maintain the cohesiveness of the particle swarm). The local deposition of chemoattractant is also subject to the same diffusion as that which affects the food sources (in fact it is represented computationally as the same `substance') and the diffusion away from the particle population also acts to generate a dynamical interface which competes with the food source gradients.

Suppression of food source gradients and local modification of gradients by the particle collective represents a highly dynamical spatial computation in which both local and distant sources of information (food source location, path availability) are integrated by both environmental and collective swarm computation. It can also be seen that the local modification of the gradient by the particle collective indirectly satisfies the second criterion for plasmodium gate construction --- attempted avoidance of local plasmodia. The dynamical gradient interface represents a fragile boundary between two separate swarms, two separate food gradients or a combination of both swarm and food gradients. The third criterion --- fusion of plasmodia can be represented in the particle model when movement of separate particle paths is limited and perturbation of the dynamic boundary occurs. This can result in fusion of network paths which corresponds to fusion of plasmodia.

\begin{figure}
\centering
\includegraphics[scale=0.8]{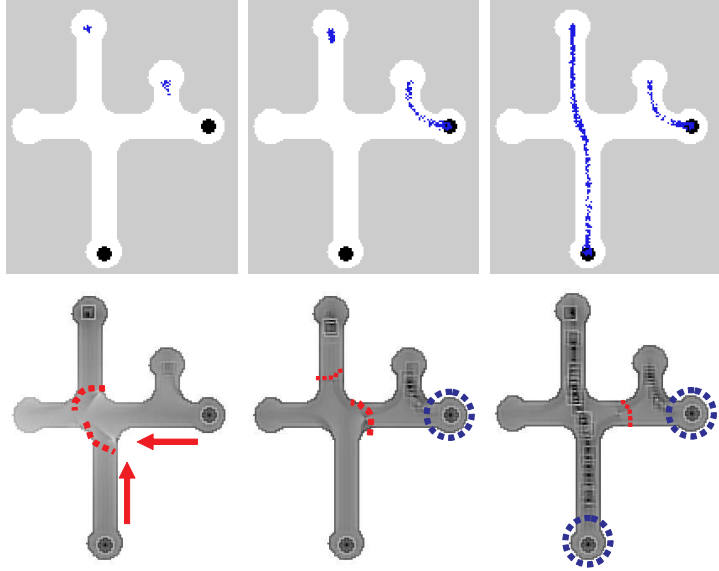}
\caption{Evolution of `plasmodium' positions and interaction fronts in the particle model for the $G_2$ gate with inputs 011.
Top Row: Particle positions.
Bottom Row: Chemoattractant gradient.
Arrows indicate propagation of gradient from food sources. 
Dashed arcs represent boundary regions separating competing gradients.
Dashed circles represent diffusion from food sources suppressed by engulfment
See text for explanation.}
\label{jeff_g2_interactions}
\end{figure}

The complex evolution of gradient fields can be seen in an example run of $G_2$ with the inputs 011 in Fig.~\ref{jeff_g2_interactions}.  The top row shows the particle positions and the bottom row shows the chemoattractant gradient field enhanced by a local method of dynamic contrast enhancement. The first column shows the propagation of chemoattractant gradient from the two food sources and the interfacial region (dashed arcs). Note that the gradient from the right suppresses the gradient from the bottom source. The second column shows the effect of suppression of the rightmost food source when engulfed by the particle population which has migrated towards it. Because the bottom food source is not suppressed the gradient from this source is stronger than the right side and the interface boundary shifts to the right of the T-junction. Note that there is also a weaker interface boundary between the diffusion gradient emanating from the bottom food source and the chemoattractant deposition from the particle population in the long vertical column. The third column shows the result of the competition between the food gradient and the population gradient --- the food gradient is stronger and the population grows and migrates downwards to the food node.

When the bottom node is suppressed the two separate paths remain stable and do not fuse. A fragile interfacial boundary can be seen between the two network paths (dashed arc) and, as long as the particles do not cross the `buffer' space between the two paths, the paths will not fuse. 

\begin{figure}
\centering
\includegraphics[scale=0.8]{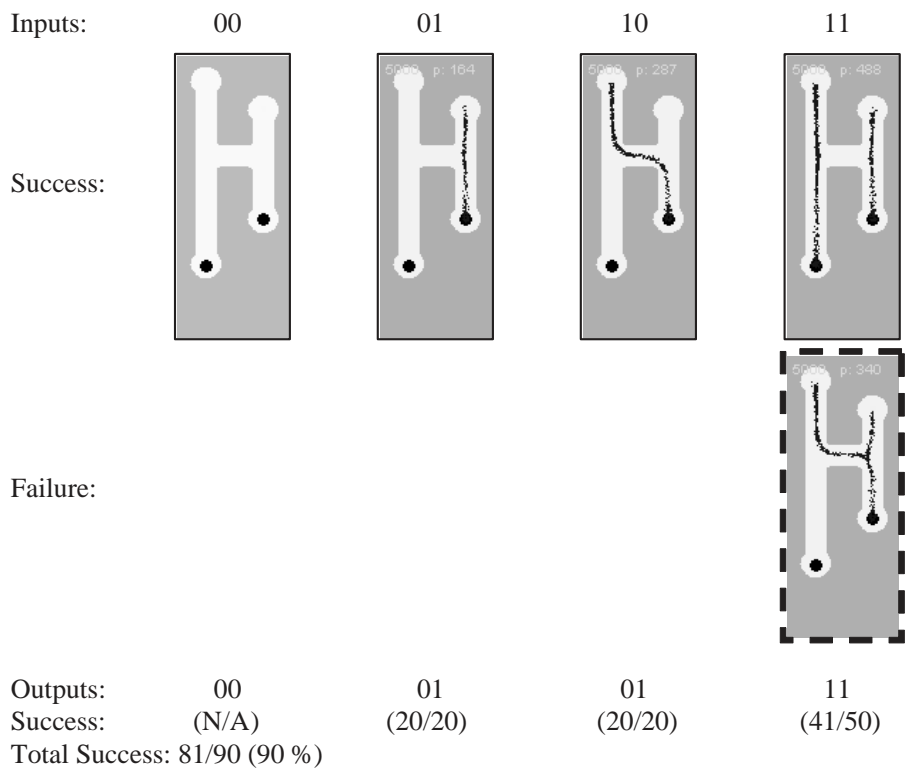}
\caption{Summary of results for particle approximation of \emph{Physarum} based logic gate $G_1$}
\label{jeff_g1_results}
\end{figure}

\begin{figure}
\centering
\includegraphics[width=0.99\textwidth]{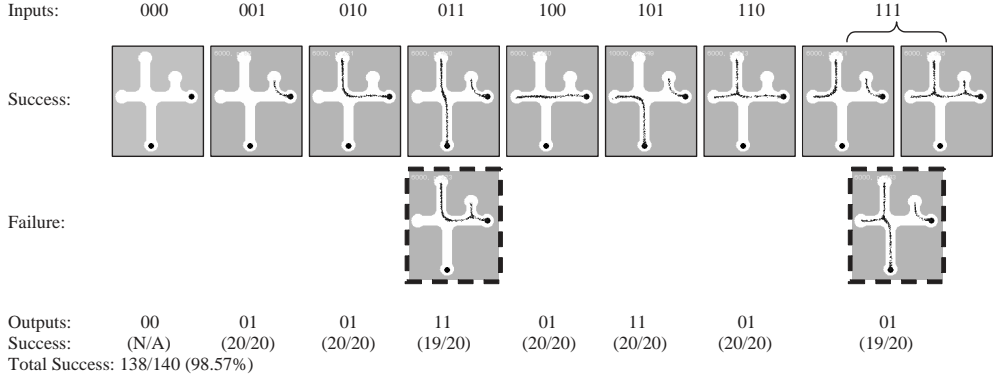}
\caption{Summary of results for particle approximation of \emph{Physarum} based logic gate $G_2$}
\label{jeff_g2_results}
\end{figure}

Results using the particle model for gates $G_1$ and $G_2$ are shown in Fig.~\ref{jeff_g1_results} and Fig.~\ref{jeff_g2_results} . The $G_1$ gate achieved 90\% reliability and the $G_2$ gate achieved 98.57\% reliability. The input conditions 0-0 were not included with the results because the output result for these inputs is guaranteed regardless of gate design. For the $G_1$ gate we see that the shorter path to the right food source attracts the simulated plasmodium in both 0-1 and 1-0 condition. Note that no branching occurs from the plasmodium to the left nutrient source when the right source is connected. This is because the movement of particles (and their deposition to the diffusion map) creates a local diffusion field around the particle collective. The strength of this locally generated field is enough to suppress the field emanating from the left food source and no branching is observed. If the strength of the local field were less than that of the nutrient source then branching and growth to the left nutrient source would indeed occur.

The errors in the $G_1$ gate all occurred in the 1-1 input condition. The `pattern' of the error is that the left particle stream did not continue downwards to the food source, but fused with the right side particle stream (indicated by dashed box). Analysis of all of the results found that whenever the growing particle plasmodium encountered a junction in a gate an apparent `hesitation' was seen. The growth tip appeared to be indecisive as to which direction to take. When a direction was eventually chosen the growth speed increased when the growth tip moved past the junction. The hesitation, and indeed some of the gate errors, was caused by disturbances in the diffusion field near the tip of the growing plasmodium. The diffusion gradient emanating from the nutrient sources is relatively uniform whereas the gradient from the plasmodium tip is more intermittent in quality (because the tip growth is non uniform and changeable in form). In contrast the gradient from a moving straight part of the particle plasmodium was more uniform. The fragility of the gradient field at the growth tip was further perturbed by the spatial changes in the environment at the junctions. This, coupled with increased possible choices of directions, led to what we describe as junctional errors. The junctional errors are characterised by failures in searching of the growing plasmodium tip and were responsible for all of the failure instances of the $G_1$ gate.

The $G_2$ gate, although more complex in design, was more reliable than $G_1$ and the only errors which occurred were a single junctional error in the 011 input condition and an error in the 111 input condition. This error was classed as a timing error and was caused by different growth rates from the two left-side inputs. Ideally the two particle streams should meet and fuse but differences in the growth of the two separate streams led to non fusion and errors in output.

To illustrate the transient dynamical nature of growth tip hesitation at junctions, junctional and timing errors, please refer to the supplementary video recordings at: (http://uncomp.uwe.ac.uk/jeff/gates.htm).

\section{Modelling the half adder}
\label{modellingadder}

\begin{figure}
\centering
\includegraphics[scale=0.6]{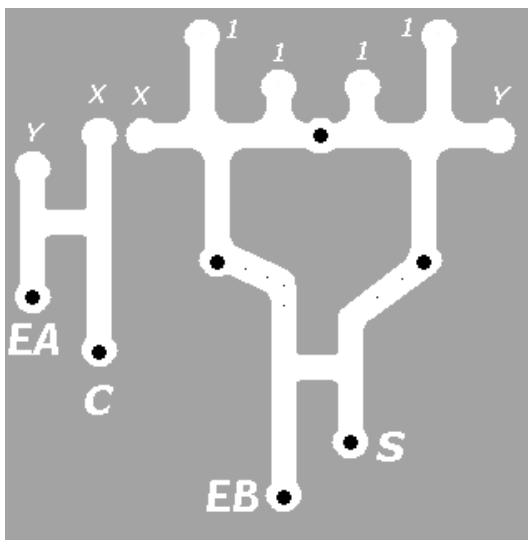}
\caption{Spatial representation of half adder based on combinations of $G_1$ and $G_2$. 
X and Y: Inputs to half adder, 1: constant {\sc Truth} signals, S: Sum output, C: Carry output.
Solid discs are food sources and small dots are small food sources to feed outputs towards lower gate inputs. EA and EB: Error checking flags (see text)}
\label{jeff_ha_design}
\end{figure}

To implement the half adder based on gates $G_1$ and $G_2$ with the particle model the scheme of the half adder in Fig.~\ref{halfadder} was slightly modified as shown in Fig.~\ref{jeff_ha_design}. The $G_2$ gate combination was simplified by `sharing' the food source between both gates. Constant {\sc Truth} inputs (`{\bf 1}') were provided as some of the gate inputs to implement the desired function. The outputs of the combined $G_2$ gates were fed to act as inputs to the lower $G_1$ gate. To ensure that the particle population continued to the input positions of the lower gate synthetic chemoattractant stimuli (small dots) were placed to guide any plasmodium along the channel to the input positions. The `$G_2$$G_2$$G_1$' triplet combination acted as the XOR (summation) part of the half adder. The {\sc and} section of the half adder (carry computation) was implemented as a single $G_1$ gate (Fig.~\ref{jeff_ha_design}, left). In the simulations the branching of initial X and Y signals to provide the inputs to both sections of the half adder was not implemented in an effort to simplify the design and the relevant X and Y inputs were introduced to the gate manually.

\begin{figure}
\centering
\includegraphics[width=0.99\textwidth]{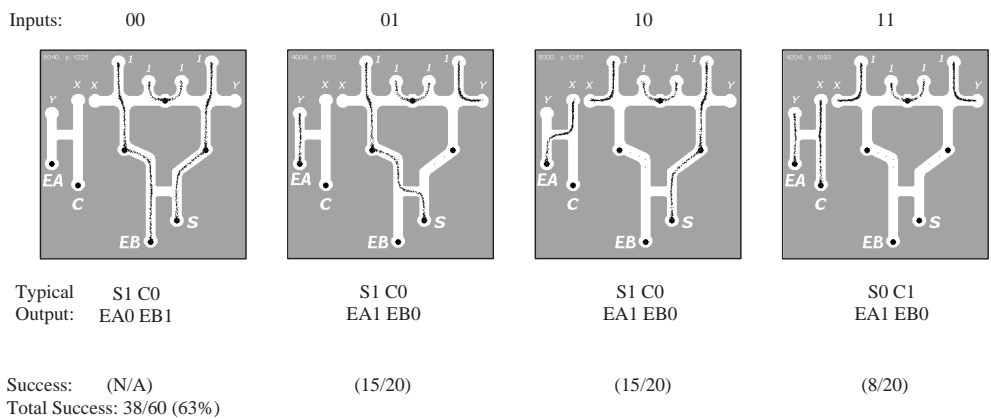}
\caption{Examples of input and output conditions for the particle approximation of the half adder}
\label{jeff_ha_results}
\end{figure}

The use of constant {\sc Truth} inputs to the half adder introduces errors in gate output when inputs are 0-0. This is because the outermost truth signals at the inputs of the $G_2$$G_2$ gates travel down through the gates and into the lower $G_1$ gate. This would result in the `no input' condition actually causing an erroneous output. Apart from redesigning the gate this presents an opportunity to consider possible use of error checking signals in the gate design. One possible error checking signal is the `EA' output in the left side of the circuit (Fig.~\ref{jeff_ha_design}, left). It can be seen that this flag should be set whenever any of the inputs are set to true. It would therefore be possible to use the absence of the EA output to indicate a 0-0 input to the half adder, and thus indicate erroneous output from the constant {\sc Truth} inputs to $G_2$$G_2$. Another possible use of outputs to indicate error conditions is the `EB' output from the left $G_1$ portion of the $G_2$$G_2$$G_1$ triplet (Fig.~\ref{jeff_ha_design}, bottom). It can be seen (Fig.~\ref{jeff_ha_results}) that the EB flag should never be set unless the 0-0 condition caused by constant {\sc Truth} inputs occurs. This flag could be combined with the lack of EA output to indicate errors. When the EB flag is set without the presence of EA then a fault can be assumed to have occurred within the half adder $G_2$$G_2$$G_1$ triplet. Of course the addition paths and mechanisms to make use of these error checking flags adds another layer of complexity to the circuitry which is out of the scope for this paper. The results of the half adder approximation can be seen in Fig.~\ref{jeff_ha_results}

The failure rate for the half adder approximation, even when not including the difficulty posed by the 0-0 configuration, was significantly higher than for the single gates. The majority of the failures were caused by timing errors, which occurred when the outermost inputs to the $G_2$$G_2$ combined gate did not fuse correctly with the constant {\sc Truth} inputs and, instead, travelled down towards the lower gate. Junctional errors also occurred three times in the left $G_1$ gate for the 1-1 input condition.

The combination and extension of the individual gates appeared to compound the errors in the individual gates. Although no definitive answer can be given as to why the unreliability increased, we speculate that the combining of the gates subtly affected the propagation and profile of the chemoattractant gradients.

\section{Discussion}
\label{discussion}

The results from the computational approximation of \emph{Physarum} support the findings of~\cite{tsuda_2004} that the organism can be used to construct simple logic gates, and also the computing schemes within this paper which explored the creation of more complex combined gates and half adder circuitry. The findings suggest that, although such circuits can indeed be built, the presence of both timing errors and junctional (search) errors would severely limit the effectiveness and practicality with even more complex circuits.

The matter of errors of the gate operations (timing errors and junctional errors) requires further consideration. The term `error' depends on the perspective taken. From an experimental viewpoint the occasionally unreliable operation of the gates is erroneous. But the notion of externally applied - by the experimenter - environment conditions and metrics of success cannot be easily applied to the behaviour of a living (or even simulated) collective organism, whose sole imperative is the location and connection of food sources for survival. By following the biological imperative the organism is not actually doing anything `wrong'. 

Ascribing such obviously successful survival strategies as erroneous behaviour merely because they do not suit the experimenter's demands seems to do a disservice to the organism. At the risk of further anthropomorphism it seems as if the collective is hampered by a \emph{double-bind} caused by conflicting biological and computational requirements: ``Please forage successfully, but not too successfully''.

It appears that this conflict could not easily be resolved from an engineering perspective because, on one hand, we cannot lessen the requirements of digital circuit operation to accept logically incorrect output. Nor is it possible to reign-in the natural foraging ability of the plasmodium. Although it is possible to track the logical errors by using spare signal channels (as noted in the half-adder circuit), acknowledging that there is an internal problem does not actually rectify the problem. Simple logic gates form the basis of complex circuits whose reliability must be total so that they may be considered as `black box' interchangeable components in hardware design. They are designed from the `top-down' to achieve their reliability and unpredictable or emergent results cannot be tolerated using this design approach. \emph{Physarum}-based computing uses the opposite approach --- simple low-level interactions generate complex and unpredictable emergent computing abilities from the `bottom-up'. Although we cannot guarantee the low-level reliability of the output of foraging in the \emph{Physarum} plasmodium we \emph{can} state that the plasmodium will forage and that the resultant emergent behaviour will be complex and unpredictable. Harnessing this spatial and temporal unpredictability, for example in a manner as suggested by~\cite{aono_neurophys}, may prove to be a more suitable application of the computational properties of \emph{Physarum}.

Although the use of \emph{Physarum} for classical logical gates indeed makes use of its intrinsic properties (gradient oriented growth, avoidance, fusion), the confinement of this naturally amorphous, dynamic and flexible organism into architectures which require precise timing and predictable path traversal does not utilise the natural advantages which the organism possesses. The increasing failure rate when circuit complexity is scaled upwards may --- perhaps fancifully --- be interpreted as the embodiment of an internal frustration by the plasmodium at such confinement and control. It appears likely that \emph{Physarum} may be more naturally suited to device implementations which harness its abilities in integrating complex, noisy, unpredictable, spatial and temporal signals. In such devices the concept of rigid control of behaviour will be reduced to an influence on behaviour, where the influence is applied as a +ve or -ve stimuli which affects behaviour. Suggested lines of inquiry lie in pattern recognition, signal generators (oscillators), signal filtration and memory storage. We plan to investigate such devices in a further study.

\section{Acknowledgment}

This research is funded by the Leverhulme Trust project F/00577/I ``Mould intelligence: biological amorphous robots''.

\end{document}